\documentclass[prb,rapid,twocolumn,superscriptaddress,showpacs]{revtex4}
\usepackage{graphicx}

\begin{document}
\title{
Scanning Fourier spectroscopy: A microwave analog study
to image transmission paths in quantum dots
}
\author{Y.-H.~Kim}
\author{M.~Barth}
\author{U.~Kuhl}
\author{H.-J.~St\"ockmann}
\affiliation{Fachbereich Physik der Philipps-Universit\"at Marburg, Renthof 5, D-35032 Marburg, Germany}
\author{J. P. Bird}
\affiliation{Department of Electrical Engineering and Center for Solid State Electronics Research,
Arizona State University, Tempe, Arizona 85287-5706, USA}
\date{\today}

\begin{abstract}
We use a microwave cavity to investigate the influence of a movable absorbing
center on the wave function of an open quantum dot. Our study shows that the absorber
acts as a position-selective probe, which may be used to suppress those wave function
states that exhibit an enhancement of their probability density near the region
where the impurity is located. For an experimental probe of this wave function selection,
we develop a technique that we refer to as scanning Fourier spectroscopy,
which allows us to identify, and map out, the structure of the classical trajectories
that are important for transmission through the cavity.
\end{abstract}

\maketitle

The ability to manipulate the wave function of mesoscopic systems offers
the potential to realize a variety of novel electronic devices,
and is also important for the study of quantum chaos,
where mesoscopic systems have been used to investigate the
signatures of wave function scarring \cite{Fro94,Fro95b,Aki97,Bir99,Mou02}.
Wave functions of a quantum-well have been measured by applying a controlled
potential perturbation method \cite{mar89,sal97}.
An important recent advance for studies in this area has been provided
by the development of scanning-probe techniques,
in particular atomic-force microscopy \cite{Top01,Top00,Woo02},
as a new experimental tool that may be used to selectively
probe the wave function of mesoscopic devices.
Investigations of this type are extremely challenging,
however, and their interpretation can be complicated by the perturbative
influence of the scanning probe on the system under study.
For this reason, it is of interest to investigate how the wave function
of a well controlled model system is influenced by the presence of an absorbing center.
It is this problem that we address in this paper, where the model system that
we consider is an open microwave cavity.
The basis for such studies is provided by the exact correspondence
between the Schr\"odinger equation for a freely moving electron in
a two-dimensional billiard, and the Helmholtz equations for electromagnetic
waves in a flat cavity \cite{Stoe99}. Due to this correspondence,
a measurement of the electric-field distribution of the cavity is equivalent
to a mapping of the billiard wave function
\cite{Stoe90,Ste95,Kuh98b,Alt95a,Dem99,Sri91,Kud95,Kim02,Blo02,Naz02,Che02},
and previous work has shown the value of such studies for the investigation
of transport in quantum dots \cite{Kim02,Blo02,Naz02,Che02}.

In this paper, we study the influence of a movable absorbing center
on the transmission properties, and the wave function, of an open microwave cavity.
The motivation for this study is provided, in part, by our recent observation
that the transmission properties of such cavities are intimately correlated
to the presence of recurring wave function scars \cite{Kim02},
similar to the behavior found previously in open quantum dots \cite{Aki97,Bir99,Mou02}.
In this paper, we show that a movable absorber may be used as a position-selective tool,
to suppress the contribution of those wave function scars that exhibit
an enhancement of their probability density close to where the impurity
is located. By employing a technique which we refer to as scanning
Fourier spectroscopy, we show how it is possible to image the electron
trajectories that dominate the transport behavior, directly from the results
of transmission measurements. We believe that the results of this work will
therefore be important for scanning-probe investigations of transport in mesoscopic dots,
and for studies of wave function scarring in open systems.

\begin{figure}
\includegraphics[width=8.6cm]{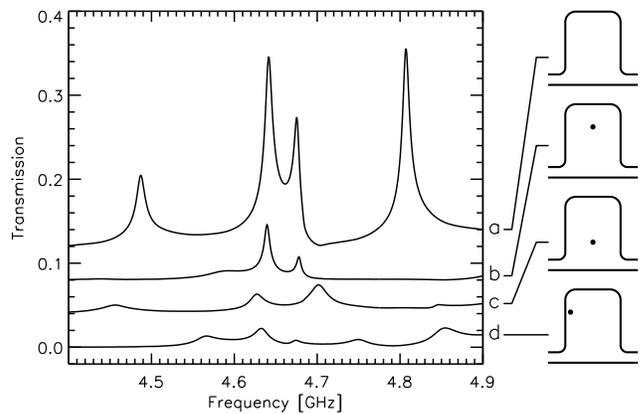}
\caption{\label{fig:1}
Microwave transmission spectrum in the regime of isolated resonances
with no absorber present (a), and for three different absorber positions
[indicated, (b)-(d)]. The spectra are shifted upwards by increments of 0.04.}
\end{figure}

The microwave cavity that we study here is the same as that described in
Ref.~\onlinecite{Kim02}. The cavity is machined from a single piece of brass,
and features waveguide leads at its bottom corners (Fig.~\ref{fig:1}) that are used
to source and detect radiation. The central cavity has in-plane dimensions
of $16 \times 21$~cm, and a uniform depth of 8~mm, ensuring that only a
single transverse mode is present at frequencies up to 18.7~GHz.
The two leads which house the antennas of the cavity are each 3~cm
wide and measurements of the microwave transmission between these antennas
are made at frequencies from 1 to 17~GHz. The upper plate of the cavity is
moved on a square grid of period 5~mm, and houses an additional antenna thus
mapping out the electric-field distribution in the plane cavity \cite{kuh00b}.
In addition, the wave functions are determined
with an absorber (diameter 1.5~cm, height 8~mm)
placed at three different fixed positions.
The reflection coefficient of the absorber is below 10\%,
i.\,e.,\ 90\% of the incoming energy is absorbed.
The diameter of the absorber has to be compared with the wavelength
of the microwaves ranging from about 30 to 1.8~cm.
In another experiment,
the antenna in the upper plate is replaced by this absorber.
Microwave transmission between the other two antennas is then measured
as the absorber is rastered back and forth across the cavity.
In other studies, the impurity is placed at a fixed position in the cavity
and the wave function is then mapped out at a series of specific frequencies.

\begin{figure}
\includegraphics[width=8.6cm]{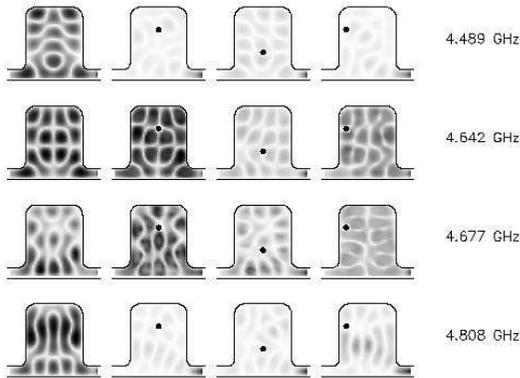}
\caption{\label{fig:2}
Wave functions for the resonances shown in Fig.~\ref{fig:1},
with the impurity located at different positions.}
\end{figure}

In Fig.~\ref{fig:1}, we show the influence of the absorber on microwave transmission through the cavity,
for a narrow frequency range where just a few isolated resonances are resolved.
It is evident from this figure that the behavior of the resonances depends sensitively on the absorber position,
and the reason for this is made apparent in Fig.~\ref{fig:2}.
Here, we show, for the clean cavity and with the impurity located at the three different positions,
the wave function at each of the resonant frequencies.
At 4.489 and 4.808~GHz, all three of these positions correspond to local
maxima in the original wave function, and the effect of the impurity on the associated resonances
is seen (Fig.~\ref{fig:1}) to be strong.
At the other two frequencies, however,
the absorber is positioned closer to nodal lines in the wave function,
and its influence on the transmission is therefore less dramatic.

\begin{figure}
\includegraphics[width=8.6cm]{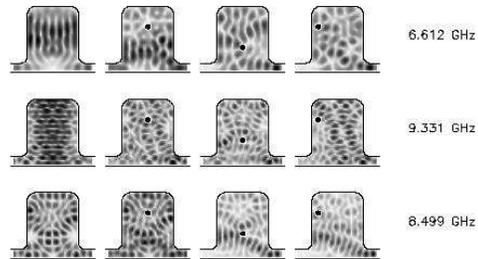}
\caption{\label{fig:3}
Wave functions for the horizontal and vertical bouncing-ball wave function,
as well as for a looplike scar. In this regime the diameter of the absorber is
comparable with the wavelength. Consequently all wave functions are destroyed by the absorber.
Only the loop wave function remains in tact, if the absorber is placed in the center of the loop.}
\end{figure}

In the left-hand column of Fig.~\ref{fig:3},
we show examples of the scarred wave functions that
were found in our earlier study of the clean cavity.
The upper panel shows the signature of a horizontal bouncing-ball orbit,
while the middle and bottom panels are scarred by vertical bouncing-ball,
and looplike, trajectories, respectively.
These scars have been shown to recur periodically as the frequency is varied,
in correspondence with the presence of distinct peaks in the Fourier spectrum
of the associated transmission fluctuations \cite{Kim02}. In Fig.~\ref{fig:3},
we see again that the absorber strongly disrupts the original wave function pattern,
whenever it is placed close to the scarred regions.
This is particularly clear in the bottom row of Fig.~\ref{fig:3},
in which we see that the impurity has little effect on the looplike scar,
when it is placed at the center of the upper loop.

In experimental studies of quantum dots, one typically measures the conductance of the system,
which is related to the quantum-mechanical transmission probability via the Landauer-B\"uttiker formula.
With this in mind, we recall that the transmission amplitude of the microwave cavity
may be written semiclassically as \cite{Gut90}
\begin{equation}
\label{eq:1}
t(k)=\sum_{n} a_n e^{ikl_n}.
\end{equation}

Here, $k$ is the wavenumber of the radiation,
$l_n$ is the length of a classical trajectory with
(complex) stability factor $a_n$, and the sum is taken
over all trajectories connecting the input and output ports.
By taking the Fourier transform of the transmission,
we can therefore obtain a stability-weighted length spectrum,
which allows a semiclassical interpretation of our results:
\begin{equation}
\label{eq:2}
\hat{t}(l)=\frac{1}{2\pi}\int t(k)e^{-ikl}dk = \sum_{n}a_n\delta(l-l_n).
\end{equation}

\begin{figure}
\includegraphics[width=8.6cm]{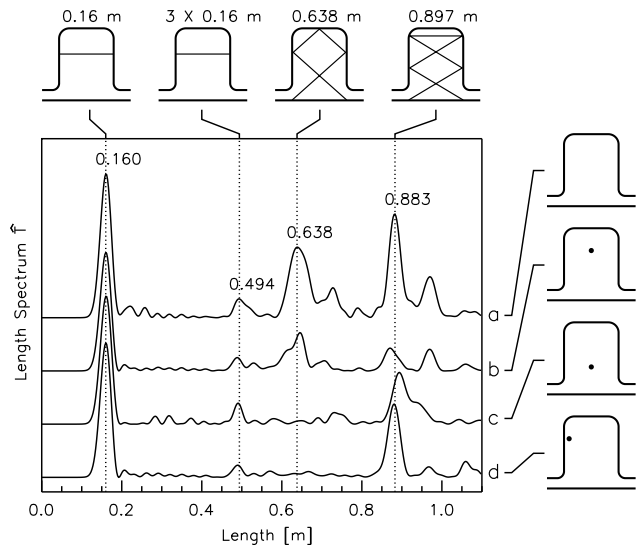}
\caption{\label{fig:4}
Length spectra $\hat{T}=|\hat{t}|^2$ obtained by a
Fourier transformation of the transmission spectrum,
of the clean cavity (a), and with the impurity located
at three different positions [indicated, (b)-(d)].
Each maximum in the length spectrum corresponds to a
trajectory connecting the entrance with the exit port.
Whenever the absorber hits a classical trajectory,
the corresponding resonance in the length spectrum is destroyed.}
\end{figure}

In Fig.~\ref{fig:4}, we show length spectra $\hat{T}=|\hat{t}|^2$,
obtained from the frequency-dependent transmission data of the microwave cavity.
With no absorber in the cavity (a), the length spectrum shows a number of peaks
which can be associated with classical trajectories,
including a harmonic of the horizontal bouncing ball.
(Since the microwave connectors produce a length length offset,
which is not accounted  for by standard calibration procedures,
the length scale was shifted by requiring that the first peak in the length spectrum
corresponds to the width of the cavity.
This was not considered explicitly in Ref.~\onlinecite{Kim02}.)
Also in Fig.~\ref{fig:4}, we indicate some classical trajectories whose lengths
are consistent with the major peaks in the Fourier spectrum.
With the absorber present in the cavity, the magnitude of these peaks
depends sensitively on its position, and our semiclassical analysis provides
an obvious interpretation of this behavior;
whenever the absorber lies close to the semiclassical trajectory associated
with a particular resonance in the length spectrum, the resonance is destroyed.
If, on the other hand, the absorber misses the trajectory, the resonance remains intact.

\begin{figure}[h]
\hspace*{-2.1cm}
\includegraphics[width=22cm]{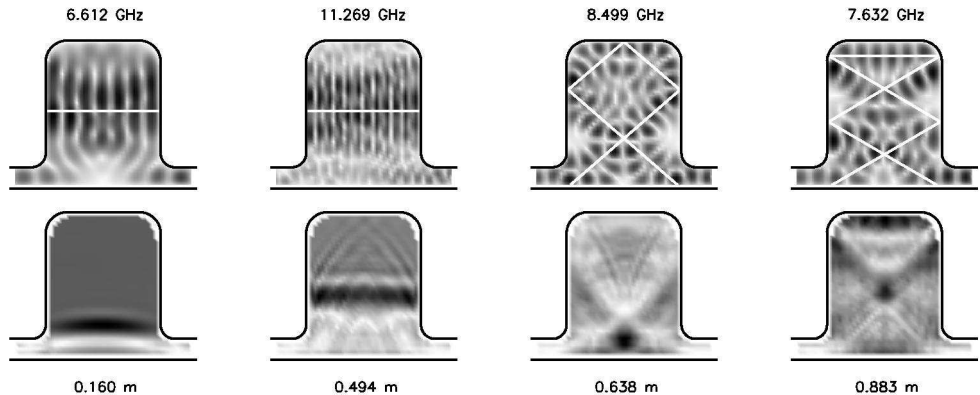}
\parbox{17.8cm}{
\caption{\label{fig:5}
The lower row shows the Fourier maps obtained for certain orbit lengths
whose corresponding trajectories are illustrated in the upper row.
In addition, a typical member of the associated scar family is depicted.
There is a close correspondence between the classical trajectories
and the light regions in the Fourier map (for further details see
the discussion in the text).}}
\end{figure}

The results of Fig.~\ref{fig:4} suggest a unique means of ``imaging"
the semiclassical trajectories responsible for transport.
In this approach, we begin by measuring the transmission
from 1 to 17~GHz, with the absorber located at each position
on the two-dimensional grid within the cavity.
At each of these positions, we then compute the length spectrum
defined in Eq.~2 and use this data to generate length-dependent Fourier maps.
These maps are constructed by plotting the variation of the Fourier power,
at a fixed orbit length, as a function of the grid coordinates.
Examples of the Fourier maps are shown in Fig.~\ref{fig:5},
where we also show the classical orbits associated with the
different scars. The grayscale in these figures indicates the
Fourier power, with dark (light) regions corresponding to minimal (maximal) Fourier power.
We see in Fig.~\ref{fig:5} that the light regions in the Fourier maps
reproduce nicely the underlying classical trajectories.
For the horizontal bouncing ball at 0.160~m, including its harmonic at 0.494~m,
the transmission takes place in the direct channel only,
therefore the absorber is effective only in this region.
Even for the quite complicated trajectories with lengths of 0.638 and 0.897~m
the orbits are reproduced by the Fourier map.
For the wave function shown at 8.499~GHz, an alternative orbit, of similar length,
hits the vertical boundaries closer to the upper corners, while at 7.632~GHz a
twin of similar length to that shown also exists.
We point out, however, that the length spectrum of Fig.~\ref{fig:4} will not resolve such pairs.
Note, however, that the spatial resolution is limited by the diameter of the absorber
and the uncertainty relation. The underlying frequency range of $\Delta \nu = 16$~GHz
corresponds to a length in space of $\Delta r = c/\Delta \nu = 1.9$~cm.

Our study suggests the possibility of mapping out the wave function
of a quantum dot in a scanning-probe experiment,
by monitoring the change in the conductance as a scanning
probe is rastered over its area \cite{Top01,Top00,Woo02}.
According to Fig.~\ref{fig:5}, such an approach should allow one
to identify the paths most relevant for transmission through the dot
(although care will be required to ensure that the perturbing tip
does not change too strongly the inherent interference conditions).
From the viewpoint of quantum chaos, the technique of
scanning Fourier spectroscopy that we have developed
may be used to obtain important information on the different electron trajectories
that dominate transmission through open dots.
While some of these trajectories connect the input and output leads
(see the trajectories with lengths of 0.638 and 0.897~m in
Figs.~\ref{fig:4} and \ref{fig:5}), as might be expected
in the usual semiclassical approximation, not all of them exhibit this property.
Note, in particular, the dominant peak in the Fourier spectrum of Fig.~\ref{fig:4},
which occurs at a length of 0.16~m, and which corresponds to a horizontal bouncing-ball trajectory.
The scar associated with this orbit is shown in Fig.~\ref{fig:5} (at 6.612 and 11.269~GHz),
and appears to buildup its probability density at regions located away from either
of the connecting leads. The role of such isolated scars has recently been considered
in the analysis of transport through open quantum dots,
where they have been shown to arise from dynamical tunnelling of electrons
through classically forbidden regions of phase space \cite{Mou02}.

In conclusion, we have used a microwave cavity to investigate the influence
of a movable absorbing center on the wave function of an open quantum dot.
Our study shows that the absorber acts as a position-selective probe,
which may be used to suppress those wave function scars that exhibit
an enhancement of their probability density near the region
where the impurity is located.
For an experimental probe of this wave function selection,
we discussed an analysis based on the technique of scanning Fourier spectroscopy,
in which we monitored how changes in the impurity position influence
the Fourier components of the transmission associated with the different scars.
This approach allowed us to identify, and map out,
the structure of the classical orbits that are important for transmission through the cavity.
We believe that our results should therefore be relevant for future scanning-probe studies
of the wave function in quantum dots.

\onecolumngrid
\twocolumngrid

Work at Marburg was supported by the Deutsche Forschungsgemeinschaft via individual grants,
while the work at ASU was supported by the Office of Naval Research
(Grant No.~00014-99-1-0326)
and the Department of Energy (Grant No.~DE-FG03-01ER45920).

\end{document}